\documentclass[twocolumn,english,prd,superscriptaddress,nofootinbib,floatfix,showpacs]{revtex4-1}

\usepackage{float,graphicx}
\usepackage[latin1]{inputenc}
\usepackage[T1]{fontenc}
\usepackage{amsmath}
\usepackage{amssymb}
\usepackage{pstricks}
\usepackage{graphicx}
\usepackage{hyperref}
\usepackage{latexsym}
\usepackage{epsfig}
\usepackage{amssymb}
\usepackage[latin1]{inputenc}
\usepackage[T1]{fontenc}
\usepackage{amsmath}
\usepackage{pstricks}
\usepackage{graphicx}

\begin{document}
\title{A Fast Route to Non-Linear Clustering Statistics in Modified Gravity Theories}
\author{Hans A. Winther}
\author{Pedro G. Ferreira}
\affiliation{Astrophysics, University of Oxford, DWB, Keble Road, Oxford, OX1 3RH, UK}

\begin{abstract}
We propose a simple and computationally fast method for performing {\it N}-body simulations for a large class of modified gravity theories with a screening mechanism such as chameleons, symmetrons and galileons. By combining the linear Klein-Gordon equation with a screening factor, calculated from analytical solutions of spherical symmetric configurations, we obtain a modified field equation whose solution is exact in the linear regime while at the same time takes screening into account on non-linear scales. The resulting modified field equation remains linear and can be solved just as quickly as the Poisson equation without any of the convergence problems that can arise  when solving the full equation. We test our method with {\it N}-body simulations and find that it compares remarkably well with full simulations well into the non-linear regime.

\end{abstract}


\maketitle


\section{Introduction}
The discovery of the accelerated expansion of the Universe is one of the biggest puzzles of modern cosmology and is attributed to an unknown substance dubbed dark energy \cite{2006IJMPD..15.1753C}.  One of the proposed solutions to this puzzle is that dark energy is a new field, with a scalar field being the simplest possibility. If such a scalar field exists and has interactions with matter, as is expected from many theories beyond the standard model, then there will be a long-rang fifth-force in nature; i.e., we have a modified theory of gravity \cite{2012PhR...513....1C}.

Results from gravity experiments on Earth (see e.g. \cite{2002cls..conf....9A}) and in the solar-system \cite{2006LRR.....9....3W} so far agree perfectly with the predictions of General Relativity (GR) and consequently any modified gravity theory must satisfy the stringent constraints coming from these experiments. This requires either that the scalar field couples to matter much more weakly than gravity or that there exists some mechanism for restoring GR in the solar-system.

Over the last decade several different types of screening mechanism have been proposed. The first class of screening is the so-called {\it chameleon} \cite{2004PhRvD..69d4026K,2004PhRvL..93q1104K} and {\it symmetron} \cite{2010PhRvL.104w1301H} mechanism. Here the scalar field is massive and the mass depends on the local matter density. If a body is screened or not depends on the value of its gravitational potential relative to a critical potential defined by the theory. A second class contains models with a shift-symmetry, $\phi \to \phi + c$, and are generally known as {\it k-Mouflage} \cite{2009IJMPD..18.2147B,2011JHEP...08..108D}. In these models screening happens for bodies that experience a large gravitational force (again with respect to a model dependent critical force). A third class contains models with a derivative shift-symmetry, $\partial\phi \to \partial\phi + c$. In this class we find the DGP model \cite{2000PhLB..485..208D,2002PhRvD..65d4023D,2003JHEP...09..029L,2004JHEP...06..059N} (in the decoupling limit) and the Galileon \cite{2009PhRvD..79f4036N,2003AnPhy.305...96A}. Screening in these models takes place for bodies that experience large force-gradients. This is the so-called {\it Vainshtein} mechanism \cite{1972PhLB...39..393V}. We should also mention screening mechanisms for models that employ a disformal coupling to matter \cite{screendisformal}. Here screening is driven by time-derivatives of the scalar field becoming small in high-density environments.

The cosmology of modified gravity models, such as the ones mentioned above, has been extensively studied. The main signature they predict, beyond modifying the background cosmology, is to alter structure formation. The scalar fifth-force present in these models is, by design, hidden in high density environments like on Earth and in the solar-system, but in the cosmological background where the density is much smaller the fifth-force can be as strong as gravity leading to potentially large signatures.

Because of this effect, to accurately study the effects on structure formation, the first line of attack is linear perturbation theory and naively one would think that on large scales this should be a good approximation. However, linear theory has the disadvantage of not taking the screening mechanism into account and it has been shown (see e.g. \cite{2012JCAP...10..002B,2013JCAP...04..029B}) for many models that linear theory gives a poor fit to the true result, found by solving the full non-linear dynamics in {\it N}-body simulations, even on scales we normally think of as linear. The reason for this is the screening effect on small scales, which represents a breakdown of the super-position principle, making the large-scale fifth-force depend sensitively on small-scale clustering. 

To calculate accurate predictions for structure formation one is therefore lead to {\it N}-body simulations. Over the last couple of years several codes have been developed to simulate modified gravity models \cite{2013arXiv1307.6748L,2012JCAP...01..051L,2013MNRAS.436..348P,2008PhRvD..78l3523O}. Such simulations need to solve the full Klein-Gordon (KG) equation for the scalar field in order to be able to calculate the fifth-force. Because the KG equation is highly non-linear this task is often hard, in terms of convergence properties, and also computational expensive. A typical {\it N}-body simulation of models in this class can easily take $10$ times as long to finish as a similar simulation for the $\Lambda$CDM model.

If modified gravity models such as those discussed above are to be confronted with observations in the non-linear regime then a fast method to compute clustering statistics would be of great value. For $\Lambda$CDM simulations, for example, such a fast method, called COLA, has recently been proposed \cite{2013JCAP...06..036T}. The goal of this paper is to investigate the possibility of a similar speed-up for modified gravity simulations, albeit with a completely different methodology: instead of trying to find a novel way of solving the exact equations we try to construct an equation that can match the behavior in regimes over which we have some analytical control, i.e. in the linear regime and the deep non-linear regime.

The field equation we propose is found by combining a screening factor, calculated from spherical symmetric configurations, with the linear Klein-Gordon equation. The screening factor depend only on the metric potential $\Phi$ which is already known to us when performing an {\it N}-body simulation (if we use a particle mesh code) and therefore does not require any additional computations. Importantly, our proposed field equation is linear (in the scalar field) which makes it simple to solve: we can use the same method as used to solve the Poisson equation for $\Phi$. The method proposal therefore has the advantage that it is able to simulate modified gravity theories taking only $\sim 1-2$ times the computational time of a corresponding $\Lambda$CDM simulation\footnote{This estimate is based on our tests on dark matter only simulations, using a particle mesh code, where the computation of the metric potential is the most time consuming part.}. The method is most suitable for particle mesh codes like $\tt{RAMSES}$ \cite{2002AA...385..337T} which we have used in this paper, but in most of our cases it should be fairly straight forward - at least in principle - to implement it in codes that do not calculate the metric potential explicitly, like for example in the popular tree code {\tt{GADGET2}} \cite{2005MNRAS.364.1105S}.

The setup of this paper is as follows: in Sec.~\ref{screensec} we briefly review the different screening mechanisms, in Sec.~\ref{appsec} we present our method for the different types of screening mechanisms, then in Sec.~\ref{testsec} we apply the method to {\it N}-body simulations before concluding in Sec.~\ref{concsec}.

\section{Screening Mechanisms}\label{screensec}
In this paper we will focus on scalar-tensor theories of modified gravity that display some sort of screening mechanism \cite{2010arXiv1011.5909K}. These are encompassed by the general (Hordenski) Lagrangian
\begin{align}\label{lagr}
\mathcal{L} = \frac{R}{2}M_{\rm Pl}^2 + \mathcal{L}(\phi,\partial\phi,\partial\partial\phi) + \mathcal{L}_m(A^2(\phi)g_{\mu\nu},\psi_m)
\end{align}
To see how screening emerges let us expand the Lagrangian about a field value $\phi_0$
\begin{align}
\mathcal{L} \simeq \frac{R}{2}M_{\rm Pl}^2 + Z^{\mu\nu}(\phi_0)\delta\phi_{,\mu}\delta\phi_{,\nu} + &\\
+m^2(\phi_0)\delta\phi + \frac{\beta(\phi_0)\rho_m}{M_{\rm Pl}} + ...
\end{align}
In a cosmological background we have $\phi_0 = \overline{\phi}$ and the scalar field produces a fifth-force on a test-mass with strength $Q \propto \beta^2(\overline{\phi})$ relative to the gravitational force. Consider now a different region of space where $\phi_0 = \phi_{\rm local} \not = \overline{\phi}$. One way to reduce the effect of the fifth-force (compared to the cosmological background) is by having a large local mass $m(\phi_{\rm local})$ which implies a very short interaction range - this is the chameleons mechanism. If the matter coupling $\beta(\phi_{\rm local})$ is small the fifth-force will also be weaker - the symmetron mechanism. Lastly, if $|Z^{\mu\nu}|$ becomes large then it leads to, after canonical normalization, a weakened matter source and therefore also a weakened fifth-force - the galileons or  k-Mouflage mechanism. There are, of course, other screening effects that cannot be understood from a simple linear expansion, but require a full non-linear analysis as we will see below.

The rough description of the different types of screening we gave above can be used to define and systematically group together different screening models.


For the purpose of this paper, we define\footnote{This is similar to what is presented in \cite{2013arXiv1312.2006K} which also offers a more detailed and pedagogical review of the different types of screening mechanisms.} three classes of screening mechanism in the following way
\begin{itemize}
\item Type I: $\phi/M_{\rm Pl} \ll \Phi_N$.
\begin{itemize}\item Type Ia: constant $\beta$ like the chameleon. \item Type Ib : field-dependent $\beta(\phi)$ like the symmetron.
\end{itemize}
\item Type II :  $|\partial\phi| \gg \mathcal{M}^2$ as in the case of kinetic theories.
\item Type III : $|\partial\partial\phi| \gg \mathcal{M}^3$ which leads to Vainshtain screening.
\end{itemize}
where $\mathcal{M}$ is some model-dependent mass-scale. The reason for this characterization is that it covers most of the known models in the literature\footnote{There can also be hybrid screening mechanisms, see for example \cite{2005PhRvD..71d3504W}. There also exists screening mechanisms for theories which has a disformal coupling to matter \cite{screendisformal}.} and also reflects the symmetries of the underlying models (which might be broken by sources). We have then that  
Type I  theories arise in the presence of a massive scalar field and have no symmetry (except a possible $\mathcal{Z}_2$ symmetry for symmetrons, Type II theories are underpinned by
shift-symmetry, $\phi\to\phi+c$ and Type III theories are associated to derivative shift-symmetry, $\partial\phi\to\partial\phi+c$.
The classification we have proposed covers a broad range of theories in the literature and we now turn to each of classes in turn.

\subsection*{Type I : The Chameleon Mechanism}
The chameleon mechanism (with the classification we use, this class also contains models like the symmetron \cite{2010PhRvL.104w1301H} and the environment dependent dilaton \cite{2010PhRvD..82f3519B}) can be found in models defined by the action
\begin{align}
S = &\int dx^4\sqrt{-g}\left[\frac{R}{2}M_{\rm Pl}^2 - \frac{1}{2}(\partial\phi)^2 - V(\phi)\right] \nonumber\\
&+ S_m(g_{\mu\nu}A^2(\phi),\psi_m)
\end{align}
where $g$ is the determinant of the metric $g_{\mu\nu}$, $M_{\rm Pl} = \frac{1}{\sqrt{8\pi G}}$ is the Planck mass, $V(\phi)$ is the self-interaction potential and $\psi_m$ are the matter fields. The Klein-Gordon equation for the scalar field becomes
\begin{align}
\square\phi + V_{\rm eff,\phi} = 0
\end{align}
In the presence of matter sources, the dynamics of $\phi$ is determined by an effective potential which (for non-relativistic matter) is given by
\begin{align}
V_{\rm eff} =  V(\phi) +  \frac{A(\phi)\rho_m}{M_{\rm Pl}}
\end{align}
For the chameleon mechanism to work there are some restrictions on the form of the potential and coupling. Roughly speaking the effective potential needs to have a minimum for any matter density $\rho_m$ and the curvature (the mass of the field) at this minimum must be an increasing function of $\rho_m$. A more thorough discussion regarding requirements on the potential and coupling can be found in \cite{2004PhRvD..70l3518B,2010PhRvD..82h3503B}.

Too see how screening works in detail we look at a static, spherically symmetric object of density $\rho_c$ and radius $R$ embedded in a background of density $\rho_\infty$. The KG equation in this case reads
\begin{align}
\frac{d}{dr}\left(r^2\frac{d\phi}{dr}\right) = r^2\left(V_{,\phi} + \frac{\beta(\phi)\rho_c(r)}{M_{\rm Pl}}\right)
\end{align}
The solution to this equation is known, in general, for two regimes. First, if the equation can be linearized then the solution gives us a fifth-force that is proportional to the gravitational force (within the Compton wavelength of the field) with strength $2\beta_\infty^2$ where $\beta_\infty = \beta(\phi_\infty)$ is the coupling strength in the background.

If the Newtonian potential of the object is much larger than some critical value, in a way made precise below, the non-linearities of the potential kick-in and the field is forced down to the minimum of the effective potential ($\phi_c$) inside the body. The exterior solution is here found to approach a critical solution \cite{2012PhRvD..86d4015B,2007PhRvD..75f3501M}
\begin{align}
\phi(r) = \phi_\infty + \frac{(\phi_c-\phi_\infty)R}{r}e^{-m_\infty r},~~~~~~r>R
\end{align}
which, remarkably, is independent of both the coupling and the mass of the body. This will be the case whenever the screening-factor
\begin{align}
\frac{\Delta R}{R} \equiv \frac{|\phi_\infty-\phi_c|}{2\beta_\infty M_{\rm Pl} \Phi_{N}} \ll 1
\end{align}
Here $\Phi_N$ is the Newtonian potential of the body and $\phi_\infty$ is the scalar-field value in the background. Note that this screening condition applies for the whole class of models and not just chameleons in particular \cite{2012PhRvD..86d4015B}.

The resulting fifth-force per unit-mass on a test-particle outside the object is given by 
\begin{align}
F_\phi &= 2\beta_\infty^2\frac{GM}{r^2}\left(\frac{\Delta R}{R}\right)(1+m_{\infty}r)e^{-m_{\infty}r}\nonumber\\
&\simeq  2\beta_\infty^2\frac{GM}{r^2}\left(\frac{\Delta R}{R}\right)~~~\text{for}~~~r \ll m_{\infty}^{-1}
\end{align}
and shows that only a small fraction
\begin{align}\label{chamscreen}
\frac{M_{\rm eff}}{M} = \frac{\Delta R}{R} \ll 1
\end{align}
of the mass of the object contributes to the fifth-force. For Type Ib models (like the symmetron) we have the additional effect that the coupling $\beta_\infty$ is field dependent and becomes small in high density regions. This again causes additional screening compared to Type Ia models.

It is also worth mentioning, as has been shown in \cite{2012PhLB..715...38B,2012PhRvD..86d4015B}, that any model in this class is also fully characterized by specifying the two 'heuristic'  functions $\{\beta(a),m(a)\}$ (instead of $V(\phi)$ and $A(\phi)$) the coupling strength relative to gravity and the mass of the field in the cosmological background\footnote{More precisely: the value at the minimum of the effective potential as a function of the scale-factor $a$. The field will generally have to follow this minimum from the early Universe and until today \cite{2012PhRvD..86d4015B}.} as a function of the scale-factor $a$. The mapping from this formulation to the formulation in terms of the potential and coupling is given by
\begin{align}\label{mbetamap}
\phi(a)  &= \phi_c + 9\Omega_mM_{\rm Pl}\int_{a_c}^a \frac{\beta(a)da}{(m(a)/H_0)^2a^4}\\
V(a) &= V_c - 27\Omega_m^2M_{\rm Pl}^2H_0^2\int_{a_c}^a \frac{\beta^2(a)da}{(m(a)/H_0)^2a^7}\\
\log A(a)  &=\log A_c + 9\Omega_m\int_{a_c}^a \frac{\beta^2(a)da}{(m(a)/H_0)^2a^4}
\end{align}
where $a_c$ is some fiducial value of the scale-factor. This formulation will be vey useful when discussing our new method below.

\subsection*{Type II : Kinetic / k-Mouflage}
This class contains models where the scalar self-interactions are governed by a kinetic function $f(X)$ and possesses a shift-symmetry in the absence of matter sources. The action is given by
\begin{align}
S = &\int dx^4\sqrt{-g}\left[\frac{R}{2}M_{\rm Pl}^2 - f(X)\right] \nonumber\\
&+ S_m(g_{\mu\nu}A^2(\phi),\psi_m)
\end{align}
where $X = \frac{(\partial\phi)^2}{2}$, $A(\phi) = e^{\frac{\beta\phi}{M_{\rm Pl}}}$ and $f(X)$ is some model specific function. A simple example is found by taking
\begin{align}
f(X) = X + \frac{1}{M^4}X^2
\end{align}
The cosmology of k-Mouflage models was recently studied in \cite{2014PhRvD..90b3507B,2014PhRvD..90b3508B}. The static spherical symmetric KG equation becomes
\begin{align}
\frac{d}{dr}\left(r^2f_X\frac{d\phi}{dr}\right)  =  \frac{\beta\rho_m(r)r^2}{M_{\rm Pl}}
\end{align}
which can be integrated up to yield
\begin{align}
f_X\frac{d\phi}{dr}  =  2\beta M_{\rm Pl}\frac{GM(r)}{r^2}
\end{align}
where $M(r) = \int 4\pi \rho_m(r) r^2 dr$ is the mass enclosed within a radius $r$. This gives a fifth-force
\begin{align}
F_\phi = \frac{GM(r)}{r^2} \times \frac{2\beta^2}{f_X(r)}
\end{align}
where $f_X(r)$ is determined from
\begin{align}
f_X^2X  &=  2 (\beta M_{\rm Pl})^2\left(\frac{GM}{r^2}\right)^2
\end{align}
This term can be written in terms of the Newtonian potential
\begin{align}\label{fxcond}
f_X^2X  &=  2 (\beta M_{\rm Pl})^2(\nabla\Phi_N)^2
\end{align}
where we have used $\frac{GM}{r^2} = \frac{d\Phi_N}{dr} = \nabla\Phi_N$. We have screening whenever $f_X \gg 1$. Unscreened objects on the other hand have\footnote{We here assume for simplicity that $f(X)$ in canonically normalized, $f(X)\to X$, in the limit $X\to 0$.} $f_X \simeq 1$. 

\subsection*{Type III : The Vainshtein Mechanism}
This class contains models with a derivative shift-symmetry in the absence of sources. The Vainsthein mechanism is responsible for the viability of massive gravity, but it can be present in other theories, most notably the Galileons. For simplicity we will here restrict our attention to the cubic Galileon model. The model is described by the action
\begin{align}
S = &\int dx^4\sqrt{-g}\left[\frac{R}{2}M_{\rm Pl}^2 - \frac{1}{2}\mathcal{L}_{\rm gal}\right] \nonumber\\
&+ S_m(A^2(\phi)g_{\mu\nu},\psi_m)
\end{align}
where $A(\phi) = e^{\frac{\beta\phi}{M_{\rm Pl}}}$ and 
\begin{align}
\mathcal{L}_{\rm gal} &= (\partial\phi)^2 + \frac{1}{\Lambda_s^3}(\partial\phi)^2\square\phi
\end{align}
Looking at a static spherical symmetric configuration we find that the KG equation becomes
\begin{align}
\frac{1}{r^2}\frac{d}{dr}\left(r^2\frac{d\phi}{dr}\right) + \frac{2}{\Lambda_s^3}\frac{d}{dr}\left(r\left(\frac{d\phi}{dr}\right)^2\right) = \frac{\beta\rho_m}{M_{\rm Pl}}
\end{align}
This equation can be integrated up to yield 
\begin{align}
\frac{d\phi}{r dr} + \frac{2}{\Lambda_s^3}\left(\frac{d\phi}{r dr}\right)^2 = 2\beta M_{\rm Pl}\frac{GM(r)}{r^3}
\end{align}
which gives that the fifth-force on a test-mass is
\begin{align}
F_{\phi} = F_N \times 2\beta^2 \times 2\left(\frac{\sqrt{1+(r_V/r)^3}-1}{(r_V/r)^3}\right)
\end{align}
where $r_V = \frac{1}{\Lambda_s}\left(\frac{2\beta M}{\pi M_{\rm Pl}}\right)^{1/3}$ is the Vainshtein radius. The fifth-force is screened whenever $r \ll r_V$. Note that we can rewrite the screening factor as
\begin{align}\label{galscfac}
\frac{M_{\rm eff}}{M} = \frac{2(\sqrt{1+ \rho(r)/\rho_{\rm crit}}-1)}{\rho(r)/\rho_{\rm crit}}
\end{align}
where $\rho(r) = \rho_m(<r) \equiv \frac{M(r)}{4\pi/3 r^3}$ is the average density within radius $r$ and $\rho_{\rm crit} = \frac{3\Lambda_s^3M_{\rm Pl}}{8\beta}$ is the critical density for screening. 

\section{An approximate method for clustering statistics}\label{appsec}
In this section we describe our method for obtaining an approximate equation to be used in {\it N}-body simulations. We will, to simplify the discussion, assume that the quasi-static approximation \cite{2014PhRvD..89b3521N}  can be applied to the models we discuss below. The quasi-static approximation states that on sub-horizon scales, time-derivatives of the scalar field (and the metric potential) can be neglected compared with spatial derivatives, i.e. we assume $|\nabla\phi| \gg H^{-1}|\dot{\phi}|$.
\\\\
The system of equations solved in {\it N}-body simulations are as follows: the {\it particle} displacement equation\footnote{Constraints on variations of constants requires that $\left|\frac{\beta\dot{\phi}}{M_{\rm Pl}}\right| \ll 2H$ so that this term can usually be neglected. This will be the case for most models considered in this paper. If this is not the case, then when applying our method we will have to use $\phi = \overline{\phi}$: the cosmological value, in the evolution equation.},
\begin{align}\label{eqgrowth}
\ddot{x} + \left(2H+\frac{\beta}{M_{\rm Pl}}\dot{\phi}\right)\dot{x} &= -\frac{1}{a^2}\left(\nabla\Phi_N + \frac{\beta}{M_{\rm Pl}}\nabla\phi\right)
\end{align}
where $\nabla\Phi_N$ is the gravitational force, $\beta \equiv \frac{d\log A}{d\phi}M_{\rm Pl}$ and $\frac{\beta}{M_{\rm Pl}}\nabla\phi$ is the fifth-force;
the {\it Newton Poisson} equation for the metric potential,
\begin{align}\label{poissoneqn}
\nabla^2\Phi_N = 4\pi G a^2\delta\rho_m
\end{align}
where we have assumed that the clustered energy density in the scalar field can be neglected (see Sec.~(\ref{secgaltest}) for more details); lastly we have the {\it modified KG} equation for the scalar field which we will discuss in what follows.

These equations form a closed system which needs to be solved every time-step of a simulation. We will derive our approximate equation below by following the procedure
\begin{itemize}
\item Derive the linearized KG equation. This follows from perturbation theory.
\item Calculate the screening factor from a static spherical symmetric configuration. Rewrite it in terms of the Newtonian potential.
\item Append the screening factor to the matter source in the linear KG equation.
\end{itemize}
In the quasi-static limit and for sub-horizon scales we quite generally find a growth equation for the matter perturbation of the form
\begin{align}\label{eq:deltam_eq}
\ddot{\delta}_m + 2H\dot{\delta}_m = \frac{3}{2}\Omega_m(a) H^2 \delta_m \frac{G_{\rm eff}(k,a)}{G}
\end{align}
where $\frac{G_{\rm eff}(k,a)}{G}$ is an effective gravitational constant that may depend on both time and scale.

Let us now focus on the modifications for each type of screening mechanism.
\subsection*{Type I}
For this class of theories, linear perturbation theory gives \cite{2012PhRvD..86d4015B} that the evolution of the matter perturbations $\delta_m$ is described by Eq.~(\ref{eq:deltam_eq}) with an effective gravitational constant
\begin{align}
 \frac{G_{\rm eff}(k,a)}{G} = 1 + \frac{2\beta^2(a) k^2}{k^2 + a^2m^2(a)}
\end{align}
Here $m(a)$ and $\beta(a)$ is the mass and coupling of the scalar field along the cosmological attractor. In real space this corresponds to the KG equation
\begin{align}
\nabla^2\phi &= a^2m^2(a)\phi + \frac{\beta(a)a^2\overline{\rho}_m}{M_{\rm Pl}}\delta_m
\end{align}
Now we recall the screening condition Eq.~(\ref{chamscreen}) for spherical symmetric configurations implies that only a fraction $\left(\frac{\Delta R}{R}\right)$ of the mass contributes to the fifth-force. To be able to interpolate between the screened regime and the unscreened regime we define
\begin{align}
\frac{\Delta R}{R} \equiv \text{Min}\left[ \frac{|\phi_\infty -\phi_c|}{2\beta_\infty M_{\rm Pl} \Phi_{N}},1\right]
\end{align}
Note that this screening condition depends only on the potential $\Phi_N$ and the scalar field-value in the background $\phi_\infty$. We propose to use the same expression with $\Phi_N$ being the metric potential ($g_{00} = -(1 + 2\Phi_N)$ in the Newtonian gauge). We replace the matter density perturbation $\delta_m$ with the effective one in the linear KG equation giving us the equation
\begin{align}\label{type1eq}
\nabla^2\phi = a^2m^2(a)\phi + \frac{\beta(a) a^2\overline{\rho}_m}{M_{\rm Pl}}\delta_m^{\rm eff}
\end{align}
with
\begin{align}
\delta_m^{\rm eff} = \delta_m \times \text{Min}\left[\frac{\phi(a)}{2\beta(a) M_{\rm Pl} |\Phi_N|},1\right]
\end{align}
We note again that $\Phi_N$ is here taken to be the metric potential in a perturbed FRLW universe and we have taken $|\phi_\infty-\phi_c| = \phi(a)$: the cosmological value. This approximation is used  since otherwise we would need to solve the full equation to get it which would render this method useless. There is however another possibility here which is to use the mapping Eq.~(\ref{mbetamap}) and replace $\phi(a) \to \phi(a(\rho_m))$: the minimum of the effective potential at a given density. If we recall that (see Eq.~(\ref{mbetamap})) any theory described by $\{V(\phi),A(\phi)\}$ can equally well be described in terms by the cosmological values of the coupling and mass $\{\beta(a),m(a)\}$, then our method allows for a direct way to perform simulations directly from a model parametrized by $\{\beta(a),m(a)\}$. 

This final equation Eq.~(\ref{type1eq}) is, as promised, linear in $\phi$, will give rise to screening in high-density environments and reduce to the linear equation on large scales. 

\subsection*{Type II}
For this class we have \cite{2007PhRvD..76b3514T} that the linear growth equation is Eq.~(\ref{eq:deltam_eq})
\begin{align}
\ddot{\delta}_m + 2H\dot{\delta}_m = \frac{3}{2}\Omega_m(a)H^2\delta_m\left(1+\frac{2\beta^2}{f_X(a)}\right).
\end{align}
Note that we have no $k$ dependence here so $G$ is modified on all (linear) scales. In real space this translates into
\begin{align}
\nabla^2\phi = \frac{\beta a^2\overline{\rho}_m}{M_{\rm Pl}f_X(a)}\delta_m
\end{align}
where $f_X(a) = f_X(X(a))$ with $X(a) = -\frac{1}{2}\dot{\phi}^2$.

In the same spirit as for Type I we now propose to include screening in this equation by appending the screening condition to the KG equation as
\begin{align}\label{fastveq}
\nabla^2 \phi = \frac{\beta a^2\overline{\rho}_m}{M_{\rm Pl}f_X(a)}\delta_m^{\rm eff}
\end{align}
where 
\begin{align}
\delta_m^{\rm eff} = \delta_m \times \text{Min}\left[\frac{1}{f_X(X(x,y,z))},1\right]
\end{align}
where $X(x,y,z) = \frac{1}{2}(\nabla\phi)^2$ is determined from $\Phi_N$ via Eq.~(\ref{fxcond}).

There is another possibility to implement screening which saves computational time. We can try to attach the screening condition directly in the force-law as
\begin{align}
\vec{F}_{\phi} &= \frac{\beta}{M_{\rm Pl}}\nabla\phi = \nabla\Phi_N \times 2\beta^2 \times \nonumber\\
& \times \frac{1}{f_X(a)} \times  \text{Min}\left[ \frac{1}{f_X(X(x,y,z))},1\right]
\end{align}
with the third factor calculated from the background solution and the last factor calculated from Eq.~(\ref{fxcond}). Note that this way of doing it is a completely different way of including the screening effect and it is not equivalent to Eq.~(\ref{fastveq}). A numerical simulations based on this procedure will be just as fast as a standard $\Lambda$CDM simulation. See the end of the next section for some importance caveats related to this procedure.

Since the screening factor for Type II depends on the force $\nabla\Phi_N$ it should also be possible to implement this in {\it N}-body codes that do not explicitly compute the gravitational potential.

No {\it N}-body simulations of these types of models exist in the literature, and such an implementation is beyond the scope of this paper, so it remains to see how accurate results this approach produces. 

\subsection*{Type III}\label{sec:type3}
In the quasi-static limit, the KG equation for the cubic Galileon model is
\begin{align}
\nabla^2\phi + \frac{1}{\Lambda_s^3 a^2}\left((\nabla^2\phi)^2 - (\nabla_i\nabla_j\phi)^2\right) = \frac{\beta a^2\overline{\rho}_m}{M_{\rm Pl}}\delta_m
\end{align}
In the linear regime we have \cite{2012PhRvD..86l4016B}
\begin{align}
G_{\rm eff} = G\left(1+2\beta^2\right)
\end{align}
which, as was the case for Type II above, gives the simple real-space equation
\begin{align}
\nabla^2\phi = \frac{\beta a^2\overline{\rho}_m}{M_{\rm Pl}}\delta_m
\end{align}
We can now attach the screening factor Eq.~(\ref{galscfac}) to the linear solution giving
\begin{align}\label{cubfast}
\nabla^2\phi = \frac{\beta a^2\overline{\rho}_m}{M_{\rm Pl}}\delta_m \times \frac{2\left(\sqrt{1+ \rho_m/\rho_{\rm crit}} - 1\right)}{\rho_m/\rho_{\rm crit}}
\end{align}
as our proposed equation. This equation again reduces to the linear one on large scales, includes screening and is linear in $\phi$. This procedure is similar to what was done in \cite{2009PhRvD..80f4023K} for the case of DGP simulations.

As for Type II we also have the possibility of attaching the screening factor directly to the force-law using the Newtonian potential
\begin{align}\label{directforcelawgal}
\vec{F}_{\phi} &= \frac{\beta}{M_{\rm Pl}}\nabla\phi = \nabla\Phi_N \times 2\beta^2 \times\nonumber\\
& \times  \frac{2\left(\sqrt{1+\rho_m/\rho_{\rm crit}} - 1\right)}{\rho_m/\rho_{\rm crit}}
\end{align}
Again we stress that this is a completely different way of including the screening effect than that described in Eq.~(\ref{cubfast}) above. As for Type II it should also be possible to implement this method in {\it N}-body codes that do not explicitly compute the gravitational potential.

There is however one serious drawback of this latter way of including the screening. The density of individual grid-cells in the simulation are very sensitive to the resolution of the simulation so if we increase the resolution the fifth-force will decrease and acctually approach zero for an infinitely resolved grid\footnote{In the limit where the grid-spacing goes to zero the density of cells that contain particles increases without bounds.}. If this method is to be used then one should therefore use a smoothed density-field like for example a top-hat with radius $R$ where the best value of $R$ would need to be fit to full simulations. Another issue related to this method is that it can violate Newton's third law: the sum of the forces on all the particles in the simulations might no longer sum to zero.

\section{Tests on {\it N}-body simulations}\label{testsec}
In this section we present tests of our method by applying it to {\it N}-body simulations. We will focus on type Ia, Ib and III mechanisms- as mentioned before, there are no fully fledged type II simulations with which we can compare our approximation. 

For simplicity we will use the same $\Lambda$CDM background cosmology and initial conditions for all the tests below\footnote{Note that the modifications we have proposed are in the gravitational sector only.}. The cosmological parameters used are $\Omega_m = 0.27$, $\Omega_\Lambda = 0.73$, $h=0.72$, $\sigma_8 = 0.8$ and $n_s=0.97$. In all the 85 simulations performed, see Table~(\ref{tab:runs}) for a list, we have used $N=256^3$ particles in a box of $B=200$ Mpc$/h$. The simulations are performed using a modified version of the $\tt{RAMSES}$ code \cite{2002AA...385..337T}. The $f(R)$ and symmetron simulations presented below have been run with 5 levels of refinements in $\tt{RAMSES}$ while the Galelion simulations have no refinements as no such code was in hand at the time the analysis was performed.

\begin{table*}
\caption{The models parameters used in the {\it N}-body test simulations. For each model we have run the full simulation, the linear simulation and our approximative simulation for a total of 85 single {\it N}-body simulations.}
\begin{tabular}{@{}lccc}
\hline\hline
Model & Parameters & Realisations & {\it N}-body implementation\\
\hline
$\Lambda$CDM	& - & 5 & \cite{2002AA...385..337T} \\	
$f(R)$~gravity	& $|f_{R0}|=10^{-4}$,~$n=1$ & 5 & \cite{2011PhRvD..83d4007Z,2013arXiv1307.6748L,2012JCAP...01..051L,2013MNRAS.436..348P,2008PhRvD..78l3524O}\\	
$f(R)$~gravity	& $|f_{R0}|=10^{-5}$,~$n=1$ & 5 & -\\	
$f(R)$~gravity	& $|f_{R0}|=10^{-6}$,~$n=1$ & 5 & -\\
Symmetron	& $\lambda_{\phi 0} = 1.0~\frac{\text{Mpc}}{h}$, $a_{\rm SSB} = 0.5$,~ $\beta=1.0$ & 5 & \cite{2012ApJ...748...61D,2012JCAP...10..002B}\\	
Cublic Galileon	 &  $c_2/c_3^{2/3}=-5.378$, $c_3=10$ & 5 & \cite{2013JCAP...10..027B}\\		
\hline\hline
\end{tabular}\label{tab:runs}
\end{table*}

\subsection*{Type Ia}

As our first test case we have chosen the Hu-Sawicky $f(R)$ model \cite{2007PhRvD..76f4004H}. {\it N}-body simulations of this model have been performed in several papers \cite{2011PhRvD..83d4007Z,2013arXiv1307.6748L,2012JCAP...01..051L,2013MNRAS.436..348P,2008PhRvD..78l3524O}, see \cite{2013arXiv1307.6748L} for the description of the code used to run the full simulations we compare our method against.

For the Hu-Sawicky model the screening factor becomes
\begin{align}
\frac{\Delta R}{R} = \frac{3}{2} \left|\frac{f_{R0}}{\Phi_N}\right| \left(\frac{\Omega_m + 4\Omega_\Lambda}{\Omega_ma^{-3} + 4\Omega_\Lambda}\right)^{n+1}
\end{align}
where $|f_{R0}|$ and $n$ are model parameters. We have performed tests for $n=1$ and $|f_{R0}| = \{10^{-4},10^{-5},10^{-6}\}$.

We perform three different simulations: i) using the full Klein-Gordon equation, ii) using the linear Klein-Gordon equation, iii) using our hybrid method. We have also performed a standard $\Lambda$CDM simulation to serve as a reference point for which we measure power-spectra and mass-functions against. The time required to perform the simulations using the new method was on average $2$ times longer, compared with $8-10$ times longer for the full simulations\footnote{The time used to run the simulation will depend sensitively on the implementation of the scalar field solver, the convergence criterion and also on the computational facilities and details. The quoted values for the full simulations are for our particular implementation of the scalar field solver so other solvers (and codes) might be able to do this faster.}, the time spend on the $\Lambda$CDM simulation for all the three models.

\begin{figure*}[htbp]
\centering
\centerline{\includegraphics[width=0.95\columnwidth]{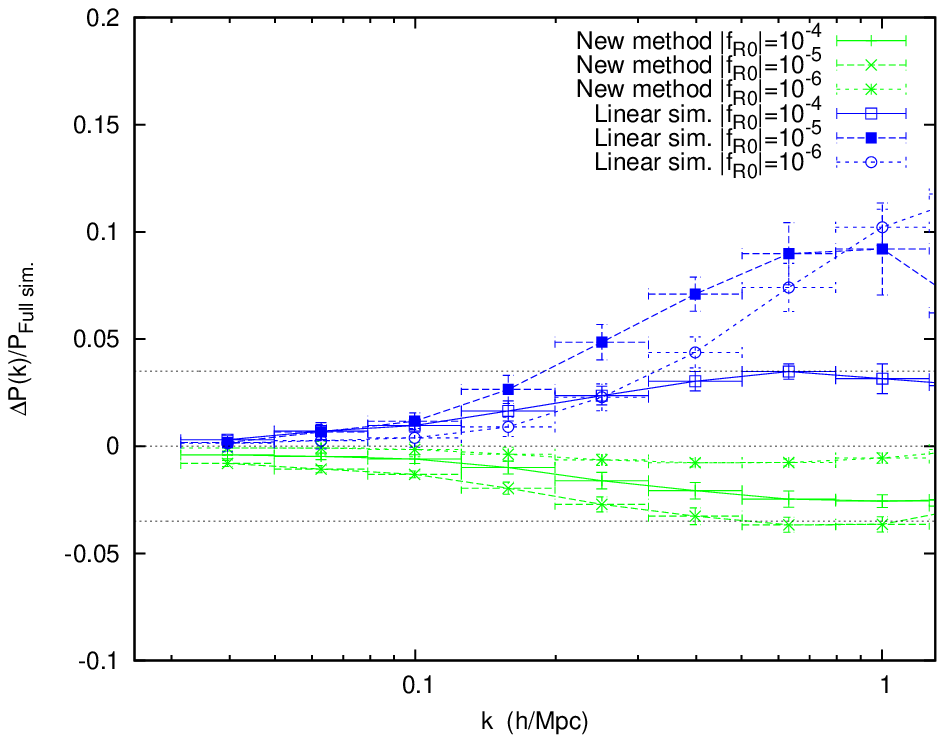}\includegraphics[width=0.95\columnwidth]{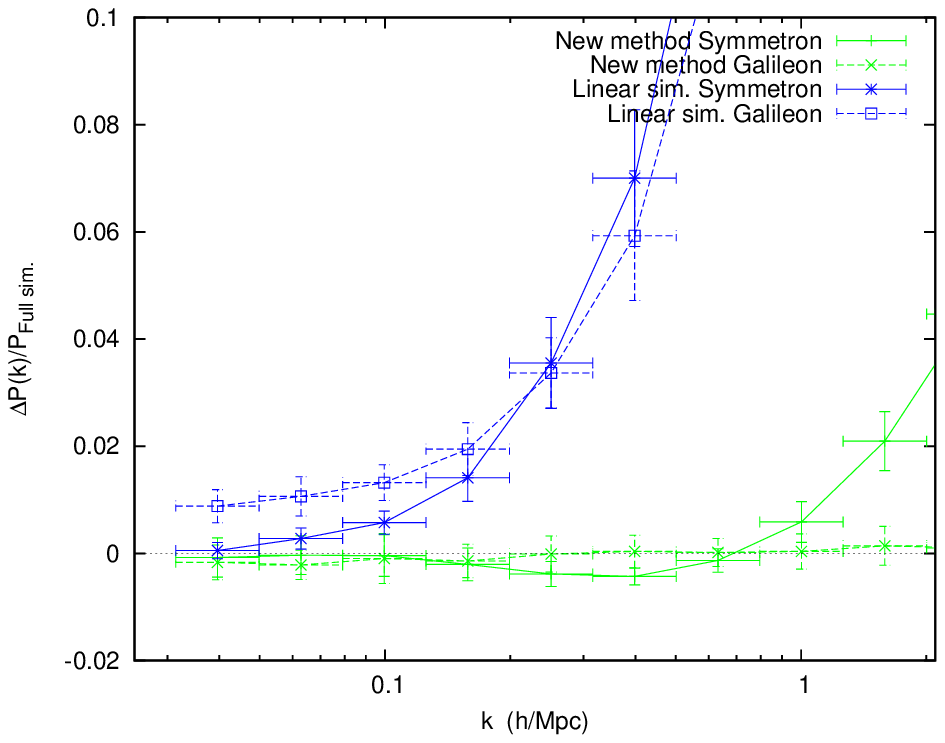}}
\caption{The fractional error in the power-spectrum with respect to the full simulations at $a=1.0$ for $f(R)$ (left) and the symmetron and Galileon (right). We show the error from simulations solving the linear KG equation in blue and our new method in green. The dashed lines in the left panel indicates $\pm 3 \%$.}
\label{fig:error}
\end{figure*}

In Fig.~(\ref{type1_fig1}) we show the fractional difference with respect to $\Lambda$CDM of the power-spectra. The agreement is very good, with errors measured with respect to the full simulations of a few $\%$ at most, see Fig.~(\ref{fig:error}). For the scales where the error reaches its maximum value the corresponding signal relative to $\Lambda$CDM is as high as 40-50 $\%$. As we go towards smaller values of $|f_{R0}|$ (which implies more screening) the agreement seems to get better and better (while worse and worse for the linear simulations). For $|f_{R0}| = 10^{-6}$ the error with respect to the full simulations is below $1\%$, compared to $\sim 10-15\%$ for the linear simulations, in the whole range of scales probed.

\begin{figure}[htbp]
\centering
\centerline{\includegraphics[width=0.95\columnwidth]{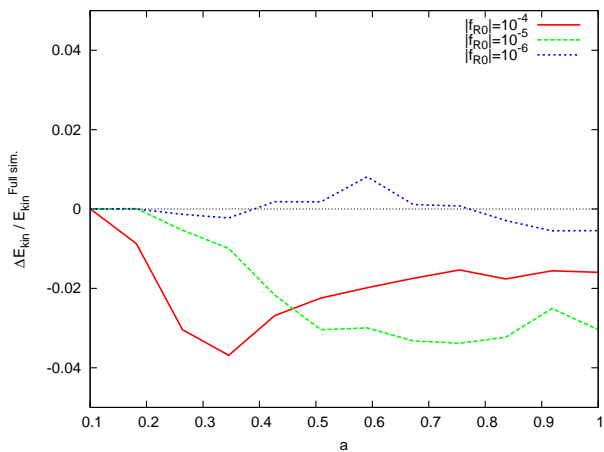}}
\caption{The fractional difference in average kinetic energy, of all the particles in our simulation box, for our approximate method with respect to the full simulations for the case of $f(R)$.}
\label{fig:errorekin}
\end{figure}

To see how well our new method is at conserving energy we have computed the time evolution of the average kinetic energy, computed from all the particles in our simulation box. Monitoring energy conservation, which is usually done by evolving the Layzer-Irvine equation, is much more involved in modified gravity simulations than for $\Lambda$CDM. A Layzer-Irvine equation for modified gravity theories was derived in \cite{2013PhRvD..88d4057W}, but it requires the scalar field $\phi$ to be computed. Since our method is designed to model the fifth-force correct and not the field value itself (which can be quite different) we choose to simply compare with the full simulations. This gives us a rough measure on how well the (global) energy is conserved and the results shown in Fig.~(\ref{fig:errorekin}) shows that the error is around $1-4\%$ throughout the evolution, i.e. of the same order of the error we find in the power-spectrum.

In Fig.~(\ref{type1_fig2}) we show the fractional difference in the mass-function with respect to $\Lambda$CDM. The agreement here is even better than for the power-spectra. The exception is the low-mass range of the $|f_{R0}|=10^{-6}$ simulations. Here we predicts too few halos for $M\lesssim 4\cdot 10^{13}~M_{\rm sun}/h$. This is likely due to our method producing too much screening in high density regions preventing additional formation of halos.

In all cases, we seem to slightly underestimate the power-spectra and the mass-function. This is a good property if we are using this approach to compare to observations then the resulting constraints will be conservative whereas if we were to derive constraints by using linear theory, which overestimates the clustering, we can rule out regions of parameter space that are in fact allowed.

\begin{figure}[htbp]
\centering
\includegraphics[width=\columnwidth]{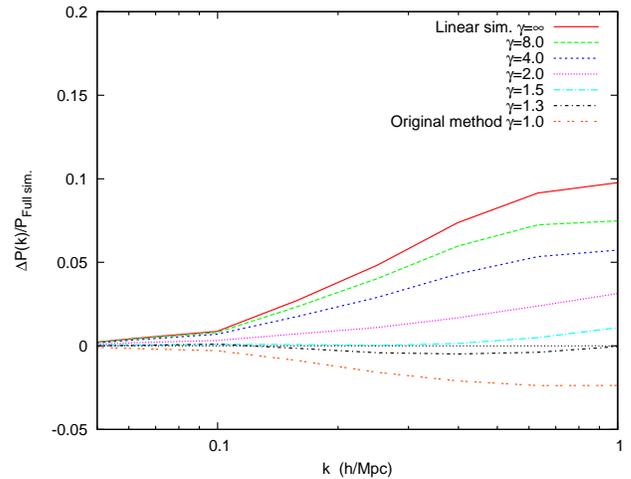}
\caption{The fractional error with respect to the full simulation for the Hu-Sawicky $f(R)$ model with $|f_{R0}|=10^{-5}$ at $a=1.0$ for different values of $\gamma$. $\gamma=1.0$ corresponds to our original method, while $\gamma = \infty$ corresponds to the linear simulation.}
\label{fig:gamma}
\end{figure}

\begin{figure*}[htbp]
\centering
\includegraphics[width=\columnwidth]{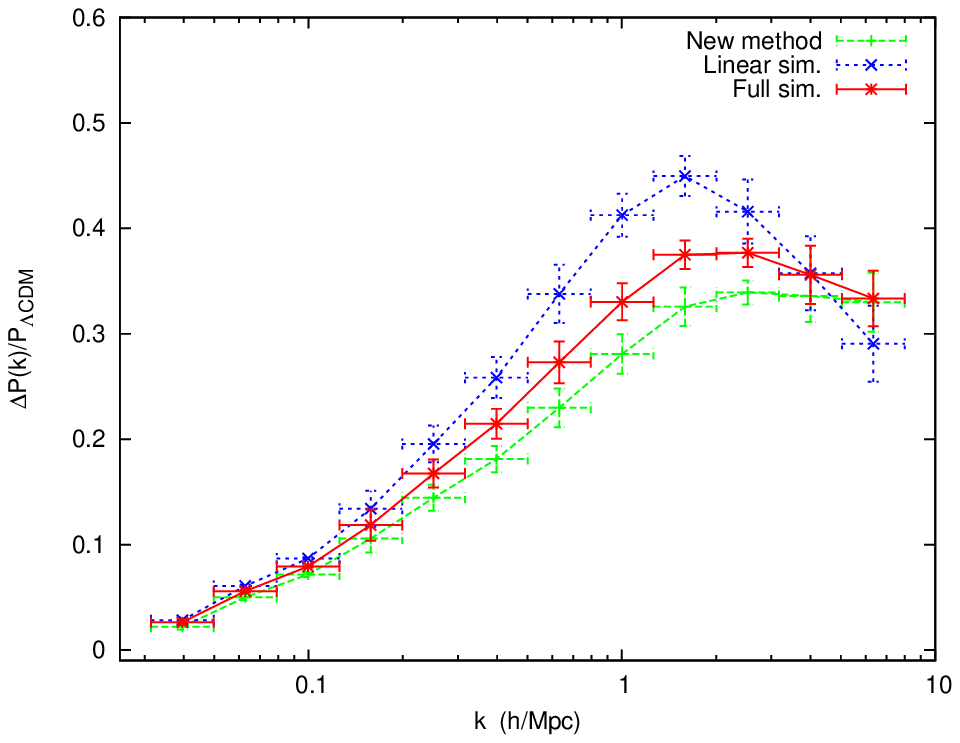}\includegraphics[width=\columnwidth]{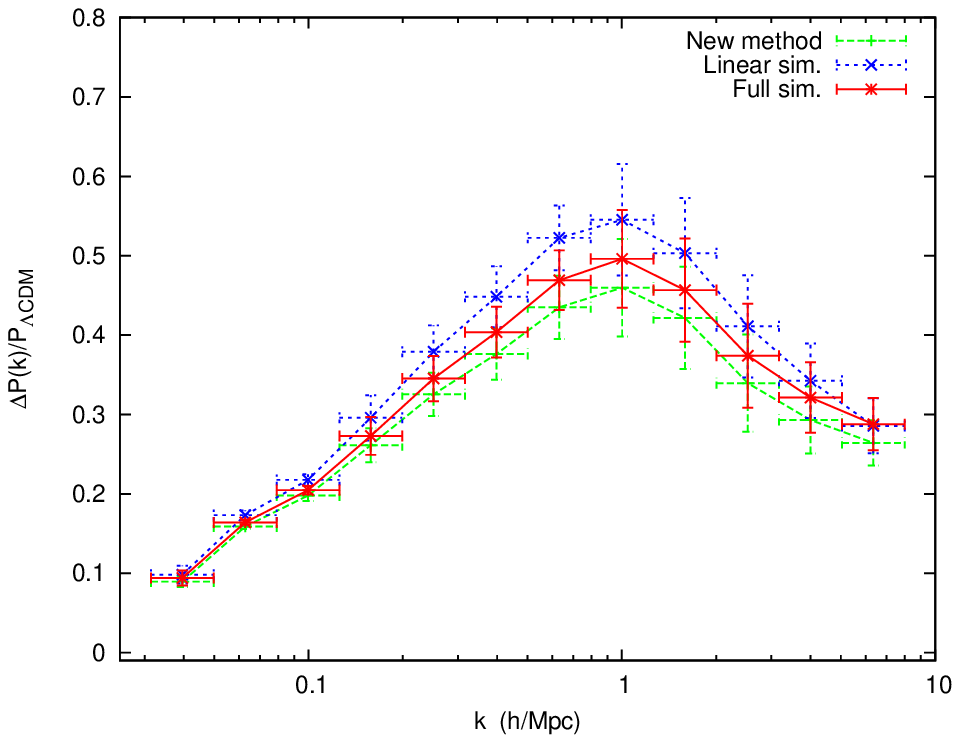}\\
\includegraphics[width=\columnwidth]{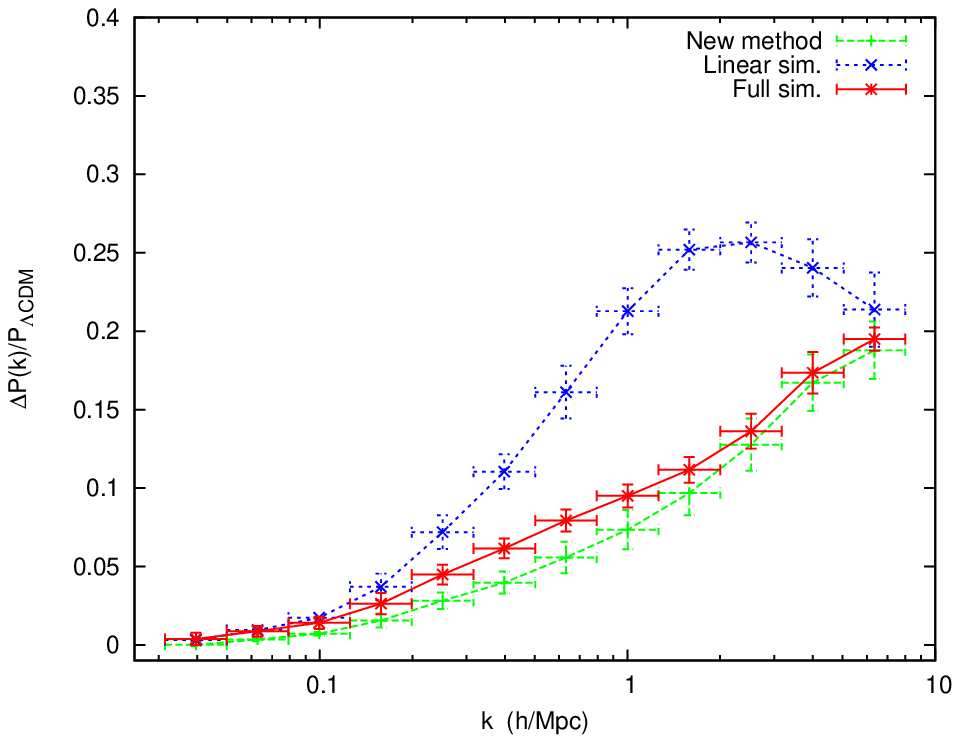}\includegraphics[width=\columnwidth]{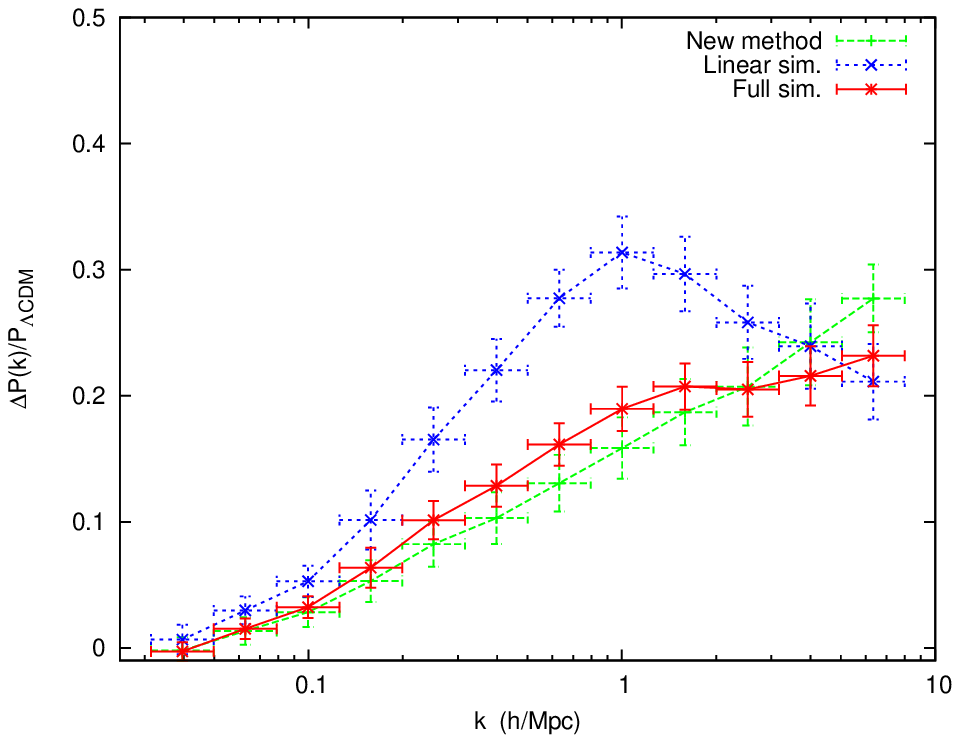}
\includegraphics[width=\columnwidth]{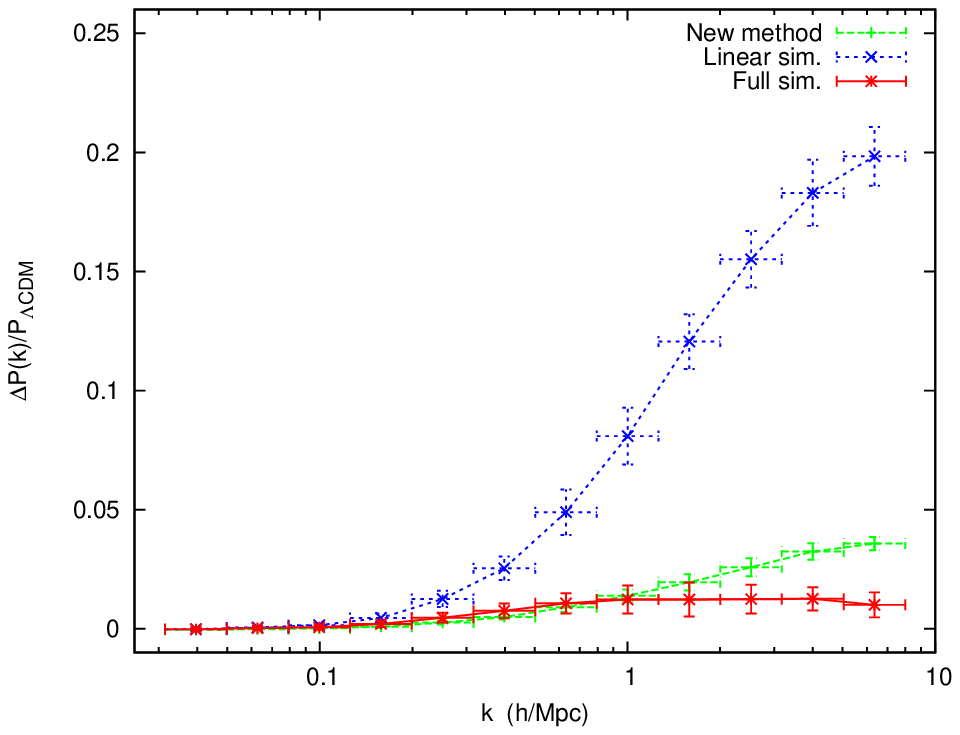}\includegraphics[width=\columnwidth]{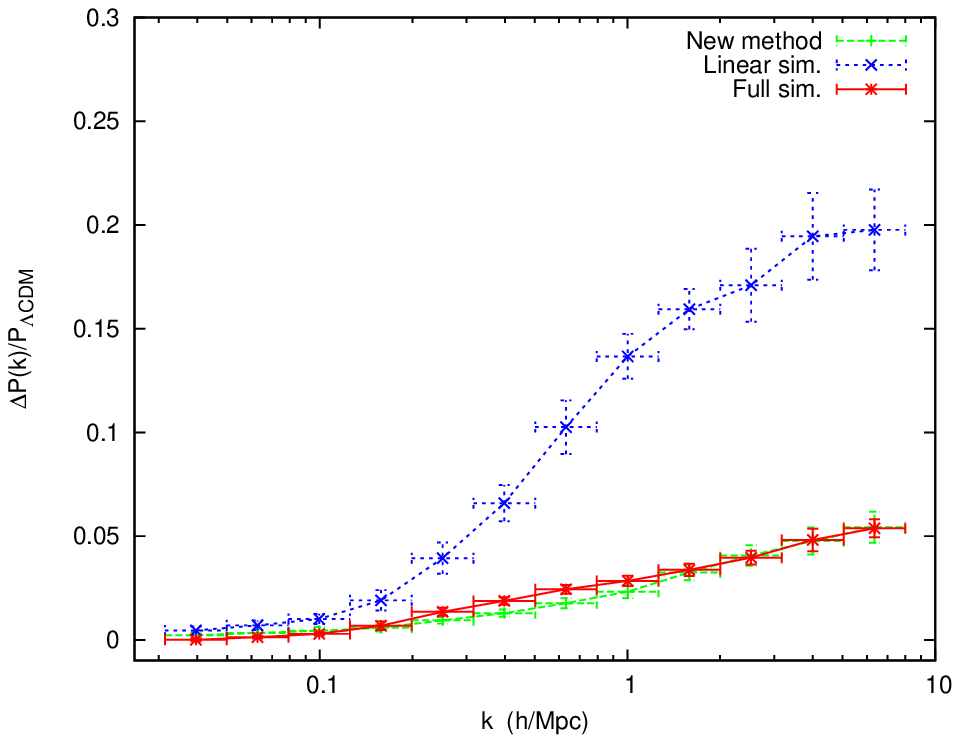}
\caption{The fractional difference in the power-spectrum for the Hu-Sawicky $f(R)$ model with respect to $\Lambda$CDM at $a=0.5$ (left) and $a=1.0$ (right). We show the results from simulations solving the full KG simulation (red), the linear KG equation (blue) and our approximate equation (green). The model parameter used are $|f_{R0}|=10^{-4}$ (top), $|f_{R0}|=10^{-5}$ (middle) and $|f_{R0}|=10^{-6}$ (bottom).}
\label{type1_fig1}
\end{figure*}

\begin{figure*}[htbp]
\centering
\includegraphics[width=\columnwidth]{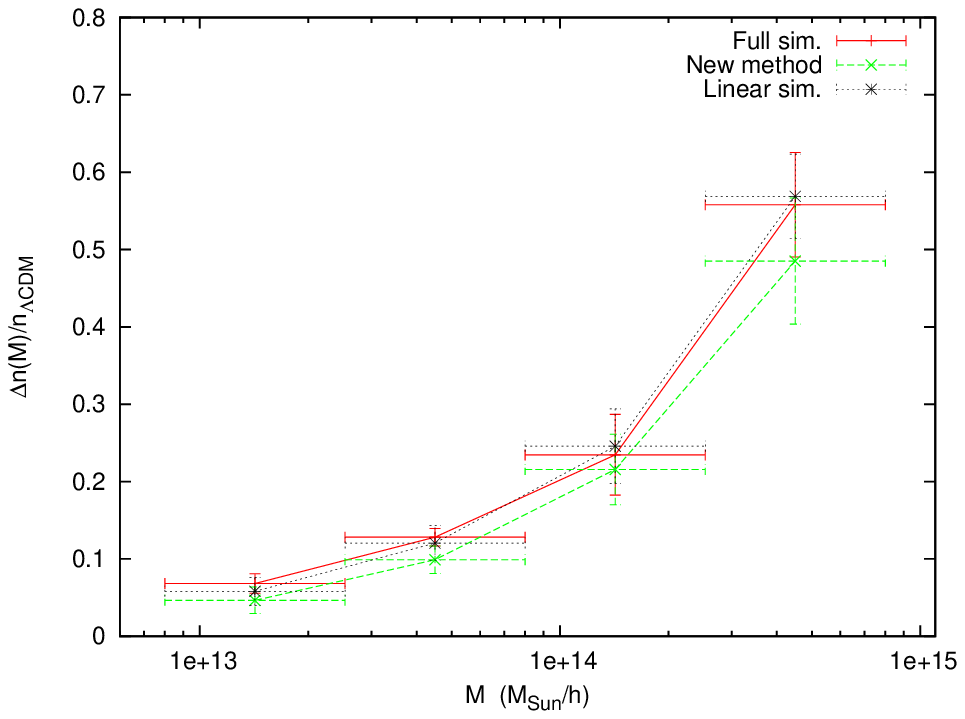}
\includegraphics[width=\columnwidth]{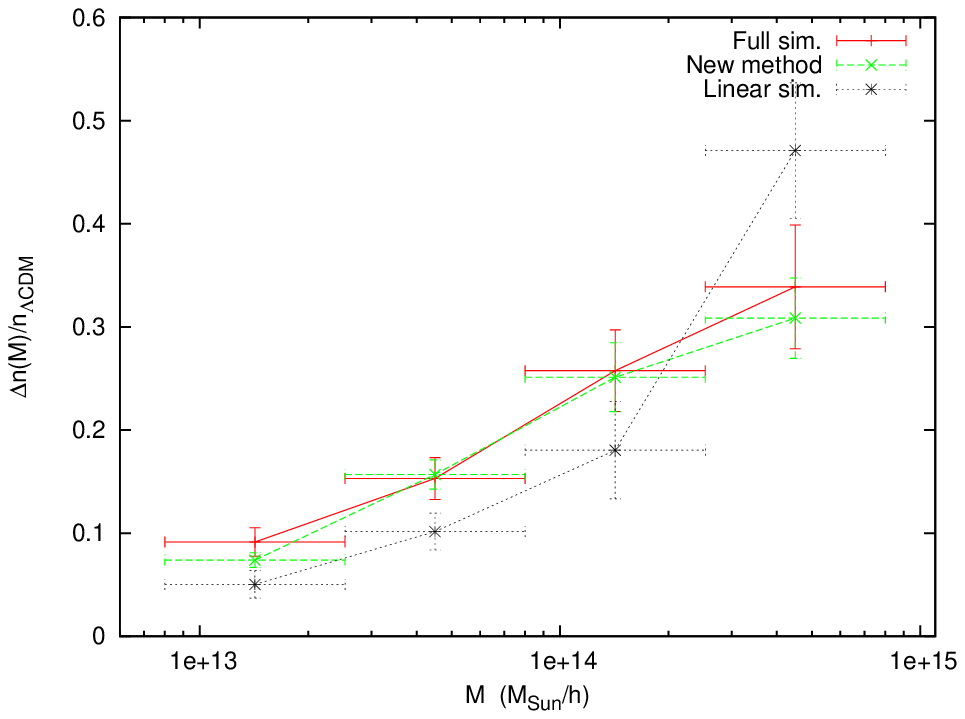}
\includegraphics[width=\columnwidth]{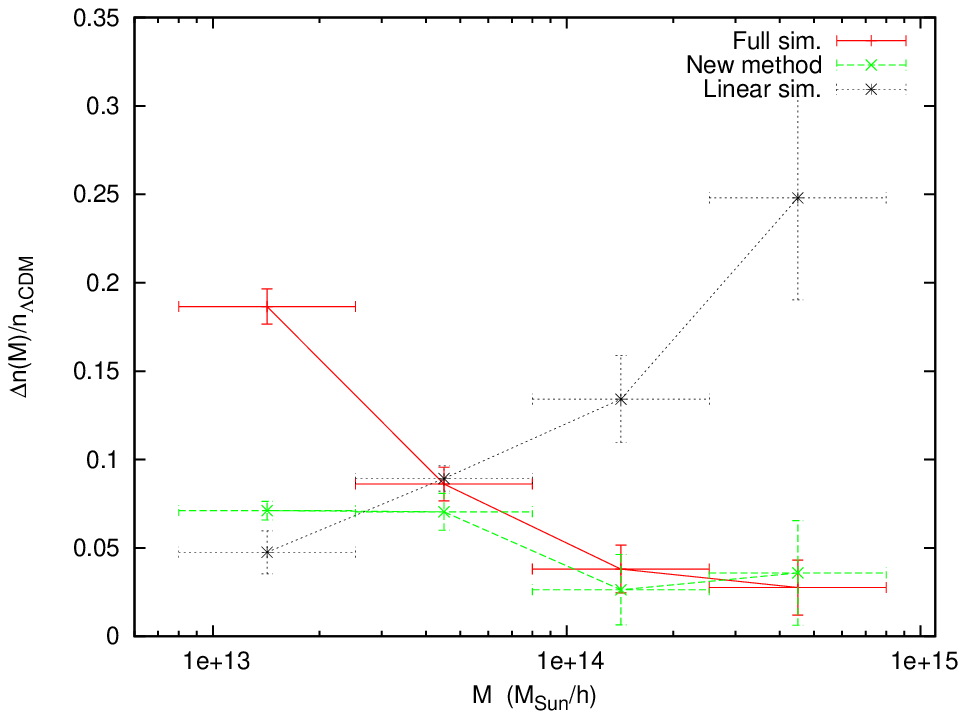}
\caption{The fractional difference in the mass-function for the Hu-Sawicky $f(R)$ model with respect to $\Lambda$CDM at $a=1.0$. We show the results from simulations solving the full KG simulation (red), the linear KG equation (black) and our approximate equation (green). The model parameter used are $|f_{R0}|=10^{-4}$ (top left), $|f_{R0}|=10^{-5}$ (top right) and $|f_{R0}|=10^{-6}$ (bottom)}
\label{type1_fig2}
\end{figure*}

One possible extension to our (zero-parameter) method, that can compensate for the slightly underestimation of power that we see, is to introduce a fudge-factor $\gamma$ as
\begin{align}
\frac{M_{\rm eff}}{M} = \text{Min}\left[\gamma \frac{\Delta R}{R} ,1\right]
\end{align}
When $\gamma \to 0$ we recover $\Lambda$CDM and when $\gamma \to \infty$ we recover the linear simulation results. Thus the effect of varying $\gamma$ is to interpolate between these two regimes. The factor $\gamma$ will have to be fitted to simulations the full results (we can do this by running a few simulations with different $\gamma$). In general we expect that $\gamma$ will have to be set to a different value for each set of model parameters used, but it might be that this factor can be set universally for each model. In Fig.~(\ref{fig:gamma}) we show the effect of varying $\gamma$ for one of the realisations of the initial conditions used in the analysis. When $\gamma=1$ we recover out original method and when $\gamma \to \infty$ we recover the linear simulation. For this particular set of model parameters we find that by taking $\gamma \approx 1.4$ we get a result that agrees with the full-simulations to $\sim 0.5\%$ accuracy for scales $k\lesssim 1~h/$Mpc.

\subsection*{Type Ib}

For our second test case in this class we have simulated the symmetron model. {\it N}-body simulations of the symmetron model have been performed in \cite{2012ApJ...748...61D,2012JCAP...10..002B}. For the symmetron model the screening factor becomes (see \cite{2012ApJ...748...61D} for the definition of the parameters)
\begin{align}
\frac{\Delta R}{R} = \frac{\Omega_m}{3.0 a_{\rm SSB}^3}\left(\frac{\lambda_{\phi 0}}{\text{Mpc}/h}\right)^2\left|\frac{10^{-6}}{\Phi_N}\right|
\end{align}
and the mass and coupling reads (for $a> a_{\rm SSB}$)
\begin{align}
m(a) &= \frac{1}{\lambda_{\phi 0}}\sqrt{1-\left(\frac{a_{\rm SSB}}{a}\right)^3}\\
\beta(a) &= \beta_0\sqrt{1-\left(\frac{a_{\rm SSB}}{a}\right)^3}
\end{align}
and $\beta = 0$ when $a< a_{\rm SSB}$. We have run simulations using the model parameters $a_{\rm SSB}=0.5$, $\beta_0=1.0$ and $\lambda_{\phi 0} = 1~$Mpc$/h$. The time spend on running our modified method was on average $2$ times slower, compared with $7$ times slower for the full simulations, compared to $\Lambda$CDM.

\begin{figure*}[htbp]
\centering
\includegraphics[width=\columnwidth]{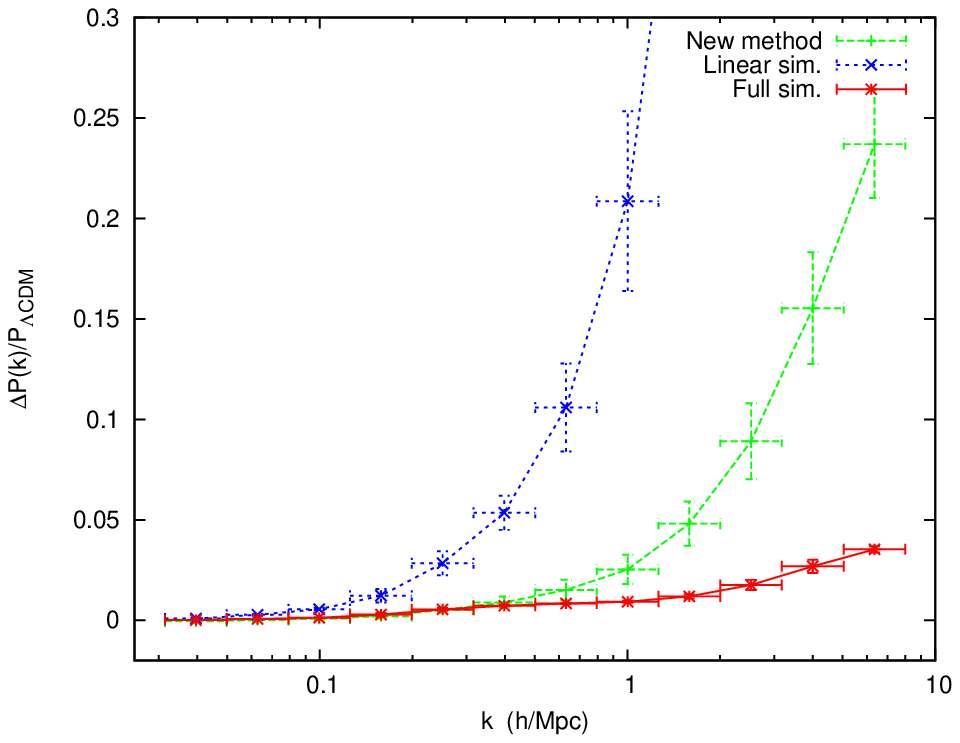}\includegraphics[width=\columnwidth]{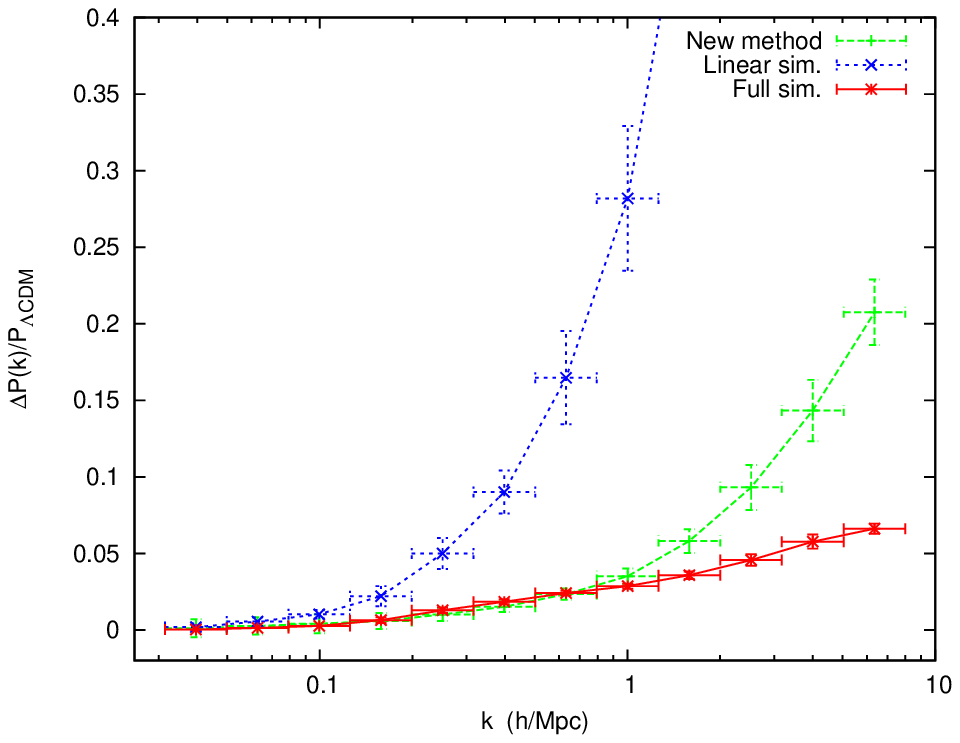}
\caption{The fractional difference in the power-spectrum for the symmetron model with respect to $\Lambda$CDM at $a=0.7$ (left) and $a=1.0$ (right). We show the results from simulations solving the full KG simulation (red), the linear KG equation (blue) and our approximate equation (green). The model parameter used are $a_{\rm SSB} = 0.5$, $\beta_0 = 1.0$ and $\lambda_{\phi 0} = 1.0~$Mpc/$h$.}
\label{type1_fig1_symm}
\end{figure*}

In Fig.~(\ref{type1_fig1_symm}) we show the fractional difference in the power-spectra with respect to $\Lambda$CDM for the full simulation and our approximate method. We see that our method is able to produce very accurate results all the way up to $k\sim 1~h/$Mpc. On smaller scales our method overestimates the clustering. A very likely reason for this is that the symmetron model has the novely, compared with $f(R)$ models, that the coupling $\beta(\phi) = \beta(a)(\phi/\overline{\phi})$ is field-dependent. In high density regions $\phi \ll \overline{\phi}$ and the coupling is suppressed. This leads to additional screening which we don't take into account.

\begin{figure}[htbp]
\centering
\centerline{\includegraphics[width=\columnwidth]{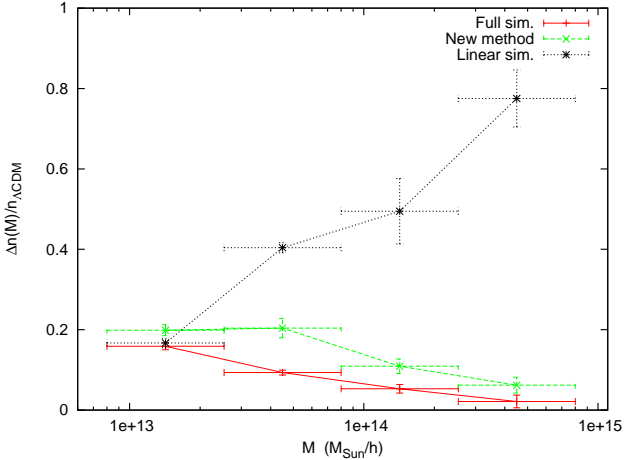}}
\caption{The fractional difference in the mass-function for the symmetron with respect to $\Lambda$CDM at $a=1.0$. We show the results from simulations solving the full KG simulation (red), the linear KG equation (black) and our approximate equation (green).}
\label{fig2_type4}
\end{figure}

In Fig.~(\ref{fig2_type4}) we show the fractional difference in the mass-function with respect to $\Lambda$CDM. Our approximate method does much better than the linear simulation, but predicts slightly more halos in the middle to high mass end. The error with respect to the true result is $\sim 5-10~\%$ (compared to $\sim~30-70~\%$ for the linear simulations). This is again likely due to our method does not take the additional screening into account.

One possible way to extend our method in this case is to make $\beta$ space dependent in the geodesic equation or in the field equation (or both). If the scalar-field tracks the minimum of the effective potential then using the $\{m(a),\beta(a)\}$ mapping we have that the value of $\beta$ in a region of space with matter density $\rho$ is given by
\begin{align}
\beta(\rho) = \beta\left(a\left(\overline{\rho}/\rho\right)^{1/3}\right) = \beta_0\sqrt{1-\frac{\rho}{\overline{\rho}}\left(\frac{a_{\rm SSB}}{a}\right)^3}
\end{align}
When the term inside the square-root is negative we have $\beta = 0$. We have implemented and tested this approach. Unfortunately, this modification was found to produce too much screening and almost no modified gravity signal was left in the power-spectrum. It might be possible to make this method viable by using a smoothed density with the smoothing radius a free parameter to be fitted by performing simulations or to introduce a fudge-factor $\gamma \leq 1$ as
\begin{align}
\beta(\rho) = \beta_0\sqrt{1-\left(1 + \frac{\gamma (\rho-\overline{\rho})}{\overline{\rho}}\right)\left(\frac{a_{\rm SSB}}{a}\right)^3}
\end{align}
or as something like
\begin{align}
\beta(\rho) = \beta_0\sqrt{1-\left(\frac{\rho}{\overline{\rho}}\right)^\gamma\left(\frac{a_{\rm SSB}}{a}\right)^3}
\end{align}
This will have the effect of reducing the screening in high density regions and could possibly be made to work. We have not tested this and leave this to future work.


\subsection*{Type III}\label{secgaltest}

As our test case we have taken the covariant cubic Galileon model \footnote{Note that the covariant Galileon simulated in \cite{2013JCAP...10..027B} is not directly coupled to matter. However, the Poisson equation for the metric potential $\Phi$ in the quasi-static limit reads
\begin{align}
\nabla^2\Phi = 4\pi G a^2\delta\rho_m + f(a)\nabla^2\phi
\end{align}
where $f$ is some time-dependent function and the particle displacement equation is simply $\ddot{x} + 2H\dot{x} = -\nabla \Phi$. By defining $\Phi_N = \Phi - f(a)\phi$ we get a standard Poisson equation
\begin{align}
\nabla^2\Phi_N = 4\pi G a^2\delta\rho_m
\end{align}
and the force-law becomes $\ddot{x} + 2H\dot{x} = -\nabla \Phi_N - f(a) \nabla\phi$. In this form the {\it N}-body equations are identical to Eqs.~(\ref{eqgrowth},\ref{poissoneqn}) and this is how we have implemented them in our code.} \cite{2009PhRvD..79h4003D}. {\it N}-body simulations of this model have been performed in \cite{2013JCAP...10..027B,2013JCAP...11..012L}. Our implementation of the full scalar field-solver is identical to that presented in \cite{2013JCAP...10..027B} (see also \cite{2013JCAP...05..023L}) and the simulations have been performed using the same best-fit parameters as found in \cite{2013JCAP...10..027B}. We have for simplicity used the same $\Lambda$CDM initial conditions as for the other simulations presented in this paper so our results are not directly comparable. We also note that our Galelion simulations have been performed using no grid-refinements in $\tt{RAMSES}$. For this test we chose to attach the screening factor directly in the force-law as described in Eq.~(\ref{directforcelawgal}). This implies we don't have to solve any scalar-field equation making the speed of the simulation very similar to that of $\Lambda$CDM.

In Fig.~(\ref{fig1_type3}) we show the fractional difference in the power-spectra with respect to our reference model which is $\Lambda$CDM simulated using the same initial conditions. The agreement is remarkable. The power-spectrum agrees perfectly for all the scales probed by the simulation. However due to the issues discussed below Eq.~(\ref{directforcelawgal}) we should be careful to conclude too strongly here. What we can conclude with is that we picked the right choice of the smoothing length for the density field (since our simulations are not refined the grid has a fixed co-moving size at all times of $R = \frac{200 \text{Mpc}/h}{256} \simeq 0.8$ Mpc/h) and that if we choose the smoothing length appropriately then our method can produce very good results. Also note that the resolution limit for the particle Nyquist frequency for the Galelion simulations is $k\sim 2~h/$Mpc so our results can not necessarily be trusted above $k\sim 1~ h/$Mpc and we have decided to cut the power-spectra here (even though the results of our method are in perfect agreement with the full simulations above this scale).

\begin{figure}[htbp]
\centering
\centerline{\includegraphics[width=0.95\columnwidth]{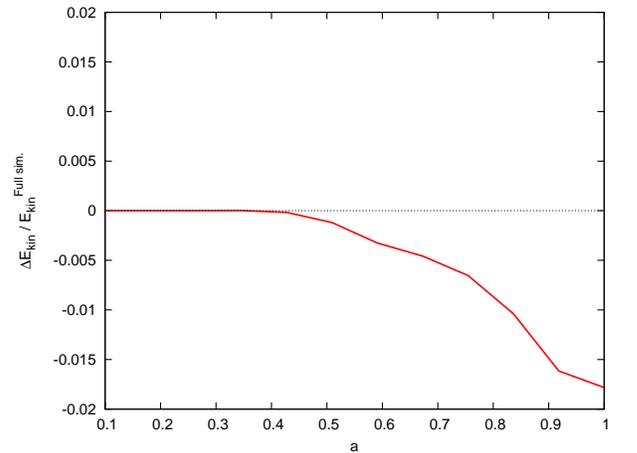}}
\caption{The fractional difference in average kinetic energy, of all the particles in our simulation box, for our approximate method with respect to the full simulations for the case of the cubic Galileon.}
\label{fig:errorekingal}
\end{figure}

\begin{figure*}[htbp]
\centering
\centerline{\includegraphics[width=0.95\columnwidth]{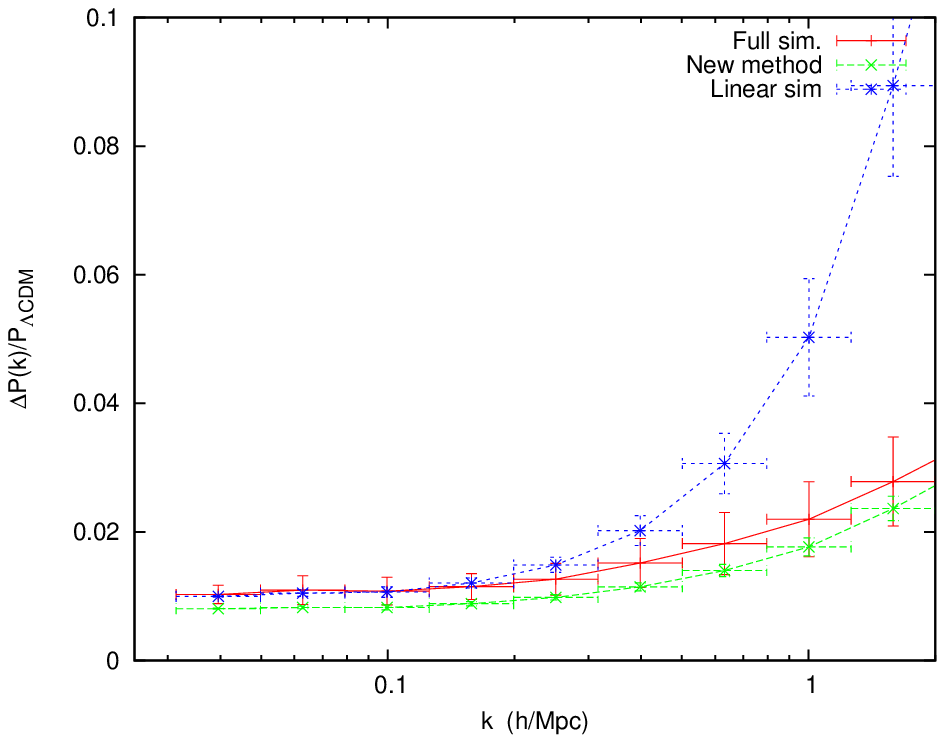}\includegraphics[width=0.95\columnwidth]{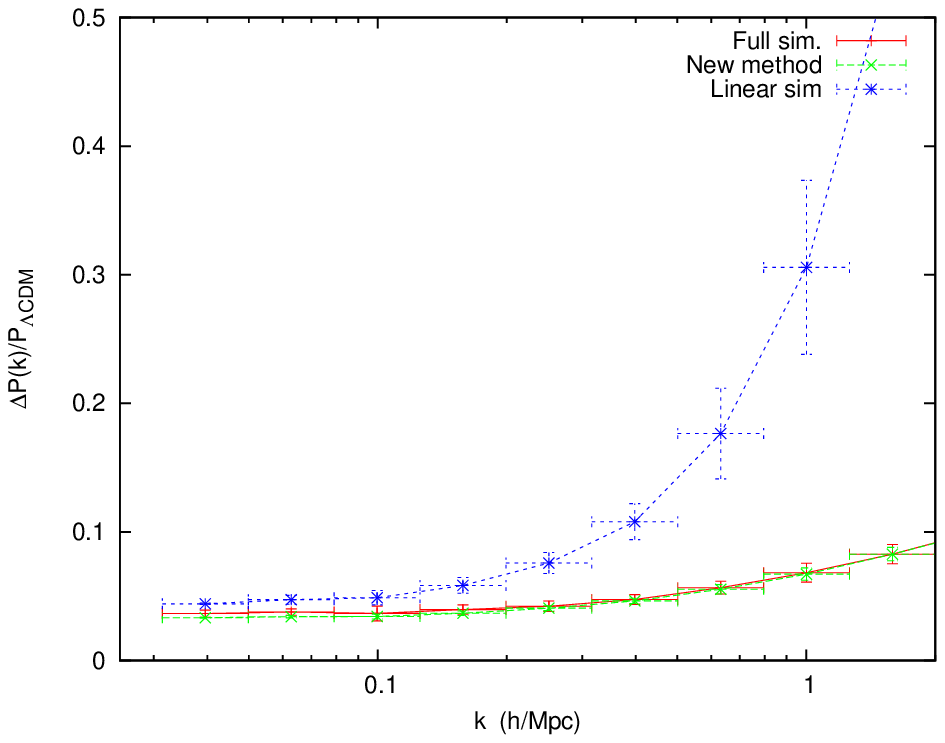}}
\caption{The fractional difference in the power-spectrum for the cubic Galileon model with respect to $\Lambda$CDM at $a=0.7$ (left) and $a=1.0$ (right). We show the results from simulations solving the full KG simulation (red), the linear KG equation (blue) and our approximate equation (green).}
\label{fig1_type3}
\end{figure*}

To see how well our method is at conserving energy (globally) we show in Fig.~(\ref{fig:errorekingal}) the evolution of the total kinetic energy of all the particles in our simulations compared with the corresponding quantity in simulations where we solve the correct equation of motion. Today we find a $2$\% deviation in the kinetic energy which is comparable with what we found for $f(R)$ gravity. We have not explicitly investigated momentum-conservation. As mentioned in Sec.~(\ref{sec:type3}) the method used do not necessarily have to respect Newton's third law, i.e. the sum of the forces on all the particles does not have to sum to zero. If this law is violated then even though it's not seen here in the power-spectrum it could be visible in higher order statistics of the density field and/or in other observables not considered here.

\begin{figure*}[htbp]
\centering
\centerline{\includegraphics[width=0.9\columnwidth]{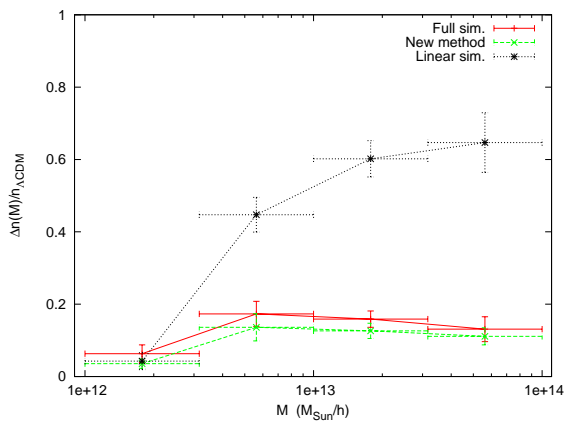}}
\caption{The fractional difference in the mass-function for the cubic Galileon model with respect to $\Lambda$CDM at $a=1.0$. We show the results from simulations solving the full KG simulation (red), the linear KG equation (black) and our approximate equation (green).}
\label{fig2_type3}
\end{figure*}

In Fig.~(\ref{fig2_type3}) we show the mass-function at $a=1.0$. Again the agreement is very good for all the halo-masses within the resolution limit. As with the $f(R)$ simulations it seems like our approach is slightly underestimating the clustering or, in other words, overestimates the amount of screening. This seems to be reasonable as in our approach the screening is local (it only looks at the local density) and this will overestimate the amount of screening at the outskirts of halos. This is, as we discussed above, a nice property as attempt to use this method to fit to observables will produce conservative constraints.

We have the opportunity to modify the screening condition (as we did for Type I above) if we need by raising the critical density for screening $\rho_{\rm crit} \to \gamma \rho_{\rm crit}$ in Eq.~(\ref{galscfac}) for some $\gamma \geq 1$ that needs to be fitted by performing simulations. For the particular model we have simulated here this does not seem to be necessary as the fit is already excellent.

\section{Conclusion}\label{concsec}

We have proposed a simple and fast, in terms of computational resources needed, method to perform {\it N}-body simulations for scalar-tensor theories which has a screening mechanism on the form described below Eq.~(\ref{lagr}).

The method consists of deriving the screening factor from studying at spherical symmetric configurations and rewriting this in terms of the Newtonian potential. This screening factor can then be attached to the linear KG equation and used in simulations.

For the three screening mechanisms studied here our method produce accurate results far into the non-linear regime, i.e. up to $k\sim \text{a~few~} h/$Mpc for $f(R)$ gravity and the Galelion and $k\sim 1~h/$Mpc for the symmetron. For the $f(R)$ models we seem to do better the further into the screening-regime we get, i.e. when the linear simulations gets further and further away from the true result. In all test cases our method seems to slightly overestimate the screening (at least on scales $k \lesssim 1~h/$Mpc). The only exception is found for the symmetron. Models where the coupling is field-dependent, such as the symmetron, can have the property that it produces additional screening on small scales. Our method, in its simplest form, does not take this into account and consequently over-estimates the power which is what we find on scales $k\gtrsim 1~h/$Mpc and the mass-function in the high-mass end. Our method can be modified to try to make the fit to the true result better by introducing a fudge-factor that parametrizes this average overestimation. 

We have only tested our method when it comes to power-spectra and mass-functions. It remains to see how good this method is at predicting other interesting observables such as halo and void profiles, halo and void shapes and velocity statistics to mention some. If attempting to apply our method, another warning is in place: the method is fundamentally phenomenological and if applied it should be tested against full simulations to get an estimate on the error. However, this only needs to be done on a few simulations compared to several tens at least needed to build up a covariance matrix.

Our method can also be useful when trying to map out the non-linear regime for a new modified gravity model not simulated before. Using our approach we can very easily implement the model, run simulations, and get a good feel for the possible signatures that it might produce.

Finally it will be very interesting to see if our method can be used in conjunction with the COLA approach \cite{2013JCAP...06..036T}  to further speed up modified gravity simulations. Such a combined method could open up the window allowing us to do a full MCMC analysis of modified gravity models using data from future large-scale structure surveys.

\section*{Acknowledgment}
We would like to thank Philippe Brax, Johannes Noller and Sigurd N{\ae}ss for useful discussion. We would also like to thank the anonymous referee for providing us with constructive comments and suggestions that helped us to improve this paper. HAW and PGF were supported by the BIPAC and the Oxford Martin School. PGF is also supported by Leverhulme and STFC. The calculations for this paper were performed on the DiRAC Facility jointly funded by STFC and the Large Facilities Capital Fund of BIS.

$\left.\right.$
$\left.\right.$

\bibliography{FastScreen_pgf}
\end{document}